\documentclass[twocolumn,superscriptaddress,nofootinbib,preprintnumbers,amsmath,amssymb,floatfix]{revtex4}
\usepackage{graphicx}
\usepackage{dcolumn}
\usepackage{bm}
\usepackage{xcolor}
\usepackage{color}
\usepackage{subfigure}
\usepackage{enumitem}

\begin{document}

\preprint{}

\title{Toward a consistent evolution of the quark-gluon plasma and heavy quarks}

\author{Marlene Nahrgang}
\affiliation{Department of Physics, Duke University, Durham, North Carolina 27708-0305, USA}

\author{J\"org Aichelin}
\affiliation{SUBATECH, UMR 6457, Universit\'e de Nantes, Ecole des Mines de Nantes,
IN2P3/CNRS. 4 rue Alfred Kastler, 44307 Nantes cedex 3, France }

\author{Pol Bernard Gossiaux}
\affiliation{SUBATECH, UMR 6457, Universit\'e de Nantes, Ecole des Mines de Nantes,
IN2P3/CNRS. 4 rue Alfred Kastler, 44307 Nantes cedex 3, France }

\author{Klaus Werner}
\affiliation{SUBATECH, UMR 6457, Universit\'e de Nantes, Ecole des Mines de Nantes,
IN2P3/CNRS. 4 rue Alfred Kastler, 44307 Nantes cedex 3, France }

\date{\today}

\begin{abstract}
Heavy-quark observables in ultrarelativistic heavy-ion collisions, like the nuclear modification factor and the elliptic flow, give insight into the mechanisms of high-momentum suppression and low-momentum thermalization of heavy quarks. Here, we present a global study of these two observables within a coupled approach of the heavy-quark propagation in a realistic fluid dynamical medium, MC@sHQ+EPOS2, and compare to experimental data from RHIC and LHC experiments. The heavy quarks scatter elastically and inelastically with the quasiparticles of the quark-gluon plasma (QGP), which are represented consistently with the underlying equation of state. We examine two scenarios: first, we interpret the lattice QCD equation of state as a sum of partonic and hadronic contributions, and second, as a gas of massive partonic quasiparticles. It is observed that independent of their momentum the energy loss of heavy quarks depends strongly on how the lattice QCD equation of state is translated into degrees of freedom of the QGP. 
\end{abstract}

% \pacs{}

\maketitle

\section{Introduction}
At top-RHIC and LHC energies a color-deconfined QCD medium of high temperatures and densities, the quark-gluon plasma (QGP), is created during ultrarelativistic heavy-ion collisions.
The properties of this fascinating new state of matter can be probed by heavy-flavor particles, which are predominantly produced in the initial hard nucleon-nucleon interactions. Due to the propagation through the colored partonic medium high-$p_T$ heavy quarks suffer from a substantial energy loss, while low-$p_T$ heavy quarks are expected to thermalize at least partially within the medium.

The nuclear modification factor, $R_{\rm AA}$, which is the ratio of the spectra measured in heavy-ion collisions to the scaled proton-proton reference, and the elliptic flow, $v_2$, which is at low-$p_T$ a measure of thermalization inside the medium and reflects at high-$p_T$  the spatial anisotropy of the initial state, are traditional observables of heavy-flavor hadrons and decay leptons.

A suppression of high-$p_T$ D mesons, heavy-flavor decay electrons and muons has been measured by the STAR \cite{StarD,Stare} and Phenix \cite{Phenixe} collaborations at RHIC and the ALICE \cite{Alice,Dainese:2012ae,Abelev:2013lca} and CMS \cite{CMS:2015hca} collaborations at LHC. The $v_2$ of $D$ mesons, heavy-flavor decay electrons and muons was found to be nonvanishing both at RHIC \cite{Adamczyk:2014yew} and at LHC \cite{Adam:2015pga}.

Perturbative QCD calculations for the average energy loss of high-$p_T$ particles include elastic \cite{elastic,Peshier:2006hi,Gossiaux:2008jv,Peigne:2008nd} and/or inelastic scatterings \cite{radiative,Wang95,Baier95,Baier97,Zakharov,GLV,Dokshitzer,AMY,ASW,Zhang:2003wk,Djordjevic:2003zk,Djordjevic:2004nq,Djordjevic:2007at,Djordjevic:2009cr}. In most of these models, no evolution of the QGP is considered and only average temperatures and path-length distributions are included. The generic form of the $R_{\rm AA}$ as a function of $p_T$ or the integrated $R_{\rm AA}$ as a function of centrality can easily be reproduced by most calculations on the basis of fundamental principles despite rather different ingredients. The strength of the suppression, however,  depends  strongly on the details of the space-time evolution of the QGP \cite{Renk:2011aa}. For quantitative predictions  the fully coupled dynamics of the heavy quarks and of the QGP needs to be taken into account.
Under the assumption that 
the time evolution of the heavy-quark distribution
function, $f(\vec{p},t)$, in the QGP can be described by a
Fokker-Planck/Langevin approach \cite{Svetitsky:1987gq,Moore:2004tg,vanHees:2004gq,vanHees:2005wb,Greco:2007sz,He:2011qa,Cao:2013ita},  
\begin{equation}
\frac{\partial f(\vec{p},t)}{\partial t}
=\frac{\partial}{\partial p_i}\left[ A_i(\vec{p})f(\vec{p},t)+
\frac{\partial}{\partial p_j}B_{ij}(\vec{p})f(\vec{p},t)\right]\, ,
\end{equation}
the interaction of a heavy quark with the QGP is expressed by a drag force $A_i$ and a diffusion tensor $B_{ij}$, which can be written as $B_\perp$ and $B_{||}$. These quantities can be calculated from the microscopic $2\rightarrow 2$ processes by 
  \begin{align}
   \frac{{\rm d}X}{{\rm d}t} =& \frac{1}{2E}
      \int \frac{d^3k}{(2\pi)^3 2 E_k}
      \int \frac{d^3k'}{(2\pi)^3 2E_{k'}}\,
      \int \frac{d^3p'}{(2\pi)^3 2E'}\, \nonumber \\
 &      \times\, \sum \frac{1}{d_i}
  \left|{\cal M}_{i,2\to2}\right|^2\, n_i(k)\,  X \nonumber \\
      &      \times\, (2\pi)^4\delta^{(4)}(p\!+\!k\!-\!p'\!-\!k')\, , 
  \label{eq:x}
  \end{align}
where $p(p')$ and $E=p_0$ ($E'=p_0')$ are momentum and energy of the heavy quark before (after) the collision and $k(k')$ and $E_k=k_0$ ($E_{k'}=k_0')$ are momenta and energies of the colliding light quark ($i=q$) or gluon ($i=g$). For the scattering process of a heavy quark with  a light quark ($qQ\to qQ$) $d_q=4$ and for the scattering off a gluon ($gQ\to gQ$) $d_g=2$. $n(k)$ is the thermal distribution of the light quarks or gluons. ${\cal M}_i$ is the matrix element for the scattering process $i$, calculated using pQCD Born matrix elements. 
In order to calculate the quantities mentioned above, $A_i$ and $B_{ij}$, one has to take $X=p-p'_i$ and $X=1/2(p-p'_i)(p-p'_j)$.
Usually, the simultaneous calculation of both coefficients from Eq. \ref{eq:x} does not satisfy the Einstein relation which assures that asymptotically $ f(\vec{p},t)$ is the distribution function at thermal equilibrium. In most Fokker-Planck/Langevin approaches one quantity is calculated and the other one is obtained via the Einstein relation under the assumption that $B_\perp= B_{||}$. It has recently been shown that the results from the Fokker-Planck/Langevin approach differ substantially from that of the Boltzmann equation in which the collision
integrals are explicitly solved \cite{Das:2013aga} because the underlying assumption, that the scattering angles and the momentum transfers are small, is not well justified. A recent review article \cite{Andronic:2015wma} gives a broad overview over the various approaches of heavy-flavor energy loss using either the Fokker-Planck/Langevin or the Boltzmann dynamics.

From Eq. (\ref{eq:x}) one sees immediately that all quantities depend on the distribution of the partonic scattering partners $n_i(k)$. In a thermal medium $n_i(k)$ is given by the Fermi-Dirac, Bose-Einstein or (if quantum statistics is neglected) the Boltzmann distribution. It is obvious that these quantities depend on the local temperature and velocities of the medium, which in most approaches are given by a fluid dynamical description of the QGP. As a consequence, final observables like $R_{\rm AA}$ and $v_2$ are strongly affected by the details of the medium evolution. While the solution of the fluid dynamical conservation equations requires only the knowledge of thermodynamic quantities, such as the equation of state, and transport coefficients, the actual nature of the quasiparticles is important for the scattering cross sections between heavy quarks and light partons. Usually, $n_i(k)$ is taken as a thermal distribution of massless, noninteracting partons \cite{Gossiaux:2008jv,Djordjevic:2007at,Djordjevic:2009cr}. The equation of state from lattice QCD calculations \cite{Borsanyi:2013bia,Bazavov:2014pvz} is not compatible with this assumption. It shows that the Stefan-Boltzmann limit is  obtained  only at extremely high temperatures which are not relevant for heavy-ion collisions.

In the present work, we address two different interpretations of the equation of state in terms of the underlying degrees of freedom.  First, we follow the phenomenological parametrization of the lattice QCD equation of state as a sum of a partonic and a hadronic contributions. Here the partons are considered as massless. This parametrization is used explicitly in  EPOS2, the fluid dynamical evolution which models the QGP in our approach. In a recent work \cite{Nahrgang:2013xaa} we investigated the effect of hadronic bound states above the transition temperatures by an adhoc parametrization of their contribution. Here, a parametrization is used, which is compatible with the underlying QCD equation of state. Second, we determine quasiparticle masses by fitting the entropy density, calculated in lattice QCD. Quasiparticle models have been used in various forms to describe the thermodynamics of QCD \cite{Peshier:2002ww,Bluhm:2006yh}. In the off-shell transport approach with a hadronic and a partonic phase, PHSD \cite{Cassing:2009vt}, a dynamical version is implemented. The approach has recently also been applied to the dynamics of charm quarks \cite{Song:2015sfa,Song:2015ykw}. Differences between our approach and the PHSD implementation include in particular the fluid dynamical versus microscopic treatment of the light parton dynamics and
the parametrization of the coupling constant which depends in our case on the momentum transfer in the collisions whereas
PHSD uses a coupling constant which depends on the temperature of the environment. 

Our model couples the Monte-Carlo treatment of the  Boltzmann equation of heavy quarks (MC@sHQ) \cite{Gossiaux:2008jv} to the $3+1$ dimensional fluid dynamical evolution of the locally thermalized QGP following the initial conditions from EPOS2 \cite{Werner:2010aa,Werner:2012xh}. EPOS2 is a multiple scattering approach which combines pQCD calculations for the hard scatterings with Gribov-Regge theory for the phenomenological, soft initial interactions. Jet components are identified and subtracted while the soft contributions are mapped to initial fluid dynamical fields. By enhancing the initial flux tube radii viscosity effects are mimicked, while the subsequent $3+1$ dimensional fluid dynamical expansion itself is ideal. Including final hadronic interactions the EPOS2 event generator has successfully described a variety of bulk and jet observables, both at RHIC and at LHC \cite{Werner:2010aa,Werner:2012xh}. 

The fluid dynamical evolution is used as a background providing us with the temperature and velocity fields necessary to sample thermal scattering partners for the heavy quarks. These scatterings can occur purely elastically or inelastically. The elastic cross sections are obtained within a pQCD inspired calculation, including a running coupling constant $\alpha_s$ \cite{Peshier:2006ah,Gossiaux:2008jv}. The contribution from the $t$-channel is regularized by a reduced Debye screening mass $\kappa m_D^2$, which is calculated self-consistently \cite{Gossiaux:2008jv,Peigne:2008nd}, yielding a gluon propagator with $1/t\to1/(t-\kappa \tilde{m}_D^2(T))$ for all momentum transfers $t$. In this HTL+semihard approach \cite{Gossiaux:2008jv}, $\kappa$ is determined such that the average energy loss is maximally insensitive to the intermediate scale between soft (with a HTL gluon propagator) and hard (with a free gluon propagator) processes. The inelastic cross sections include both, the incoherent gluon radiation \cite{Gunion:1981qs}, which has been extended to finite quark masses \cite{Aichelin:2013mra}, and the effect of coherence, i.e. the Landau-Migdal-Pomeranchuk (LPM) effect \cite{Gossiaux:2012cv}. 
The spatial diffusion coefficient from this approach \cite{Berrehrah:2014tva,Ozvenchuk:2014rpa} is compatible with the available lattice QCD calculations \cite{Borsanyi:2010cj}, which currently have still large uncertainties. In order to further constrain the model, we rescale the cross sections by a global factor $K$, which is chosen such that the results give a reasonable agreement with the $R_{\rm AA}$ data at intermediate and high $p_T$. For $\sqrt{s_{\rm NN}}=2.76$~TeV at LHC, $K_{\rm c}^{\rm LHC}=1.5$ for purely collisional processes and $K_{\rm c+r}^{\rm LHC}=0.8$ for collisional+radiative(LPM) processes. The rescaling is less well determined at the $\sqrt{s_{\rm NN}}=200$~GeV RHIC energy, because high-$p_T$ data is not yet available. In this work we apply the same $K$-factors at RHIC as at LHC.
All other observables are then calculated with the same rescaled cross sections. Results presented within this model so far \cite{Nahrgang:2013xaa,Nahrgang:2013saa,Nahrgang:2014vza} assume massless thermal partons.

This paper is organized as follows. 
We describe the two approaches to the coupling between the heavy-quark sector and the fluid dynamical medium via the equation of state in section \ref{sec:eos}. In section \ref{sec:drag} we investigate the consequences of this coupling on the drag coefficient before we present the results for the $R_{\rm AA}$ and $v_2$ of the full evolution at LHC and RHIC energies in section \ref{sec:results} and the conclusions in section \ref{sec:conclusions}.

\section{Equation of state and coupling to the heavy quarks}\label{sec:eos}
Thanks to the continued progress in lattice QCD calculations at vanishing net-baryon density, today the QCD equation of state is known very precisely \cite{Borsanyi:2013bia,Bazavov:2014pvz}. This knowledge tremendously reduced uncertainties in fluid dynamical simulations of ultrarelativistic heavy-ion collisions and facilitated quantitative estimates for the value of the shear viscosity as well as constraints for the initial state.
For heavy-flavor dynamics the use of a realistic space-time evolution, i.e. of an approach that reproduces the available data on bulk observables, is therefore possible and should become standard for reliable quantitative statements about heavy-flavor energy loss and thermalization in heavy-ion collisions.

At this level of precision the question of the nature of the active degrees of freedom in the QGP arises, as it affects the matrix elements as well as the thermal distribution function of partons in Eq. (\ref{eq:x}). 
In  \cite{Nahrgang:2013xaa} we discussed the reduction of heavy-flavor energy loss in the presence of color-neutral hadronic bound states above $T_c$ as was advocated in \cite{Ratti:2011au}. For an estimated fraction of hadronic bound states, the $R_{\rm AA}$ at LHC could still be reproduced by increasing the $K$-factor for a collisional plus radiative scenario from $K_{\rm c+r}^{\rm LHC}=0.8$ (for a $100$~\% partonic medium) to $K_{\rm c+r}^{\rm LHC}=1.0$. It was found that the $v_2$ was more sensitive to the smaller fraction of partons in the medium at later times of the evolution. According to some lattice calculations for the quark-number susceptibilities there are indications for the existence of hadronic bound states above $T_c$, although this is excluded by other investigations \cite{Bazavov:2014xya} on the lattice. While a definite statement or even a quantitative description is not yet available, one might refer to model studies \cite{Steinheimer:2010ib, Turko:2011gw} which all give an estimate for the fraction of hadronic bound states by either adjusting their thermodynamic quantities to the ones calculated on the lattice or according to the model ingredients. None of these models is, however, able to reproduce the lattice (off-diagonal) quark-number susceptibilities correctly, which motivated the work in \cite{Ratti:2011au}, and which are the essential quantities when the existence of hadronic bound states above $T_c$ is claimed. It is thus not clear if these models capture the proper physics of hadronic bound states around the transition temperature.

Here, we follow two other approaches. The first is closely related to the approach of  \cite{Nahrgang:2013xaa} but here
we treat the equation of state as implemented in the EPOS approach. There the lattice equation of state is parametrized, above $T_f = 134.74$~MeV, as a sum of an ideal partonic gas and of a hadron resonance gas (HRG). Assuming that heavy quarks interact only with the colored partonic medium the heavy quark energy loss will be reduced as compared to a model in which is assumed that above $T_c$ only partons exist. For the second approach, no coexistence of partons and hadrons is assumed, but the partons  constituting the medium above a temperature $T_f<T_c$ are massive quasiparticles. These masses can be determined as a function of the temperature by fitting the equation of state.

\subsection{EPOS parametrization of the lattice equation of state}
There are several parametrizations of the lattice QCD equation of state that connect a high-temperature part to a hadronic medium at lower temperatures \cite{Huovinen:2009yb,Bluhm:2013yga}. The parametrization used in the EPOS2 fluid dynamical simulations relies on an effective hadronic and an effective partonic contribution such that the pressure reads
\begin{equation}
p(T)=p_{\rm QGP}(T)+\tilde\lambda(T)(p_{\rm HRG}-p_{\rm QGP})\, ,
\label{eq:pressurefit}
\end{equation}
where $p_{\rm QGP}$ is the pressure in the Stefan-Boltzmann (SB) limit of QCD, i.~e. the pressure of an ideal ultrarelativistic plasma of massless quarks and gluons
\begin{equation}
p_{\rm QGP}=\frac{d_g+7/8 d_q}{90}\pi^2T^4\, .
\end{equation}
The degeneracy factors of the gluons and quarks are $d_g=2\times(N_c^2-1)$ and $d_q=2_{\rm spin}\times2_{q\bar{q}}\times N_c\times N_f$ with $N_c=3$ colors and $N_f=3$ flavors. 
The hadronic contribution is given by the pressure of the hadron resonance gas model
\begin{multline}
 p_{\rm HRG}/T^4=\frac{1}{VT^3}\sum_{i\in {\rm mesons}}\ln{\cal Z}_{m_i}^M(T,V,\mu_B,\mu_Q,\mu_S)\\
                 +\frac{1}{VT^3}\sum_{i\in {\rm baryons}}\ln{\cal Z}_{m_i}^B(T,V,\mu_B,\mu_Q,\mu_S)\, ,
\end{multline}
and
\begin{equation}
 \ln{\cal Z}_{m_i}^{M/B}=\mp\frac{Vd_i}{2\pi^2}\int_o^\infty{\rm d}k K^2\ln(1\mp z_i\exp(-\epsilon_i/T))\, ,
\end{equation}
with the energies $\epsilon_i=\sqrt{k^2+m_i^2}$ and the fugacities
\begin{equation}
 z_i=\exp((B_i\mu_B+Q_i\mu_Q+S_i\mu_s)/T)\, .
\end{equation}
$\mu_X$ are the chemical potentials for baryon number ($X=B$), electric charge ($X=Q$) and strangeness ($X=S$). 
The EPOS parametrization assumes that below the temperature $T_f=134.74$~MeV the equation of state is given by a pure HRG and thus $\tilde{\lambda}(T<T_f)=1$.
Above $T_f$ $\tilde\lambda(T)$ is obtained from a fit of (\ref{eq:pressurefit}) to lattice calculations \cite{Borsanyi:2010cj} and with following form
\begin{multline}
\tilde{\lambda}(T)=\exp\bigg[-\left(\frac{T-T_f}{\delta(1+(T-T_f)/(\delta a))}\right)\\
                     -b\left(\frac{T-T_f}{\delta(1+(T-T_f)/(\delta a))}\right)^2\bigg]\, .
\end{multline}
As shown in \cite{Werner:2012xh} this parametrization reproduces well the lattice equation of state at $\mu_B=0$ for the following parameter: $\delta=0.24$~GeV, $a=0.77$, $b=3.0$.

\begin{figure}
 \centering
 \includegraphics[width=0.45\textwidth]{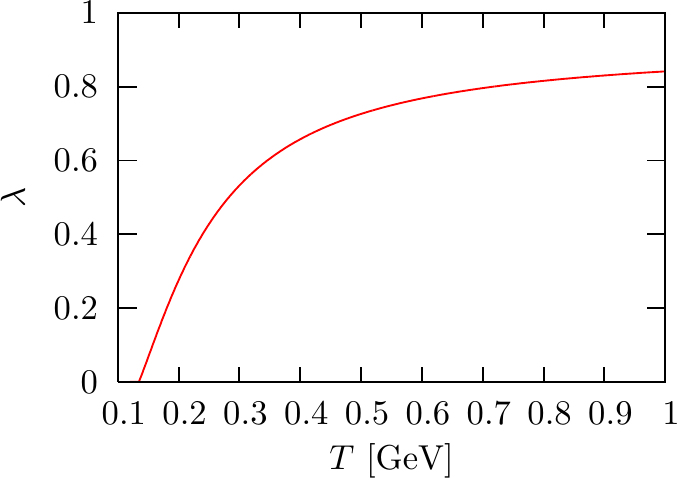}
 \caption{(Color online) The fraction of partonic degrees of freedom as given by the EPOS parametrization of the lattice equation of state.}
 \label{fig:lambdaepos}
\end{figure}

The fraction of the effective partonic degrees of freedom, given by $\lambda=1-\tilde{\lambda}$ in equation (\ref{eq:pressurefit}), is shown in Fig. \ref{fig:lambdaepos} which should be compared to 
the equivalent figure of \cite{Nahrgang:2013xaa}. We observe that $\lambda$ approaches unity very slowly. Even at temperatures as high as $T\sim 1$~GeV, which can be reached locally in the initial hot spots at LHC energies,  it is still $\lambda\approx0.84$. In reality the resulting difference can, however, not stem from hadronic contributions, as the present parametrization suggests.
We assume that the difference to the  Stefan-Boltzmann limit is due to some residual interactions at high temperatures, which effectively reduce the possibility for the heavy quarks to scatter off  constituents of the colored medium.

\subsection{Thermal masses of the light quarks}

Our second approach is an  extension of  the model established in \cite{Gossiaux:2008jv} by assuming that the incoming and outgoing 
light partons, which interact with the heavy quarks, have a finite mass. For this purpose, we treat those as long-living quasiparticles.
It is well known that quasiparticle models are able to reproduce the lattice QCD equation of state \cite{Bluhm:2004xn, Cassing:2009vt,Berrehrah:2013mua} by assuming effective dispersion relations for noninteracting quasi-quarks and -gluons in the QGP. Due to the statistical factor of $\exp[-m/T]$ we expect that in a medium with a given temperature the density of light massive partons is reduced as compared to the density of massless partons, what leads to a reduced scattering rate.

\begin{figure}[tb]
 \centering
 \includegraphics[width=0.45\textwidth]{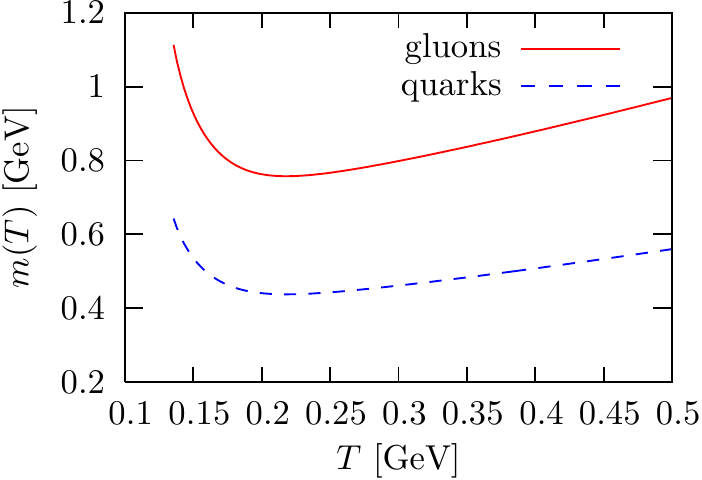}
 \caption{(Color online) Thermal masses of the quarks and gluons in the QGP within a quasiparticle approach.}
 \label{fig:thmass}
\end{figure}

The temperature dependence of the parton masses is obtained from fitting the entropy density of a noninteracting gas of massive quarks and gluons to the lattice equation of state \cite{Borsanyi:2013bia,Bazavov:2014pvz}.

The pressure and the energy density read
\begin{multline}
p(T)=d_q\int\frac{{\rm d}^3p}{(2\pi)^3}\frac{p^2}{3 E_q}f_{\rm FD}(E_q) \\
     + d_g\int\frac{{\rm d}^3p}{(2\pi)^3}\frac{p^2}{3 E_g}f_{\rm BE}(E_g)-B(T)
\label{eq:pthmass}
\end{multline}
\begin{multline}
e(T)=d_q\int\frac{{\rm d}^3p}{(2\pi)^3} E_q f_{\rm FD}(E_q) \\
     + d_g\int\frac{{\rm d}^3p}{(2\pi)^3} E_g f_{\rm BE}(E_g)+B(T)
\label{eq:ethmass}
\end{multline}
with $E_q=\sqrt{p^2+m_q^2}$, $E_g=\sqrt{p^2+m_g^2}$ and the temperature dependent bag constant $B(T)$. $f_{\rm FD}$ and $f_{\rm BE}$ are the Fermi-Dirac and Bose-Einstein distributions respectively. In order to connect $m_q$ and $m_g$ we use the perturbative HTL-result $m_g=\sqrt{3} m_q$ \cite{Thoma:1995ju} as a conservative estimate. We assume the same thermal masses for $u$, $d$ and $s$ quarks.
The mean-field contribution $B$ cancels in the entropy density
\begin{equation}
s(T)=\frac{e(T)+p(T)}{T}\, .
\label{eq:entropyfit}
\end{equation}

 The thermal masses of quarks and gluons are shown in Fig. \ref{fig:thmass}. At high temperatures we find the almost linear behavior as it is known from pQCD calculations. The quasiparticle masses show a strong increase for temperatures above and close to $T=134$~MeV, which coincides very well with $T_f$ from the EPOS parametrization. In this simple quasiparticle picture no assumption about the structural form of the temperature dependence of the thermal masses is made. Other quasiparticle approaches \cite{Peshier:2002ww,Bluhm:2006yh} express the masses via the perturbative form $m^2\propto g^2T^2$ and parametrize a logarithmic temperature-dependence of the coupling $g$ by a fit to the lattice QCD equation of state. It is thus assumed that the nonperturbative physics in the vicinity of the transition temperature can effectively be described by a temperature-dependent coupling that strongly increases near $T_c$. The definition of the running coupling constant at finite temperatures is not unique. In our approach we do not assume any explicit temperature dependence of $\alpha_s$. The coupling is determined by the momentum transfer in the individual scattering process. This is different to the PHSD approach in which the coupling constant is a function of the  temperature. It shows a strong enhancement near $T_c$ leading there to an increase of $A_i$ and hence of the heavy-flavor energy loss \cite{phsdHF} compared to what we expect in our approach (although in our approach there is as well an effective temperature dependence because at a  smaller temperature the momentum transfer is smaller and therefore the coupling is larger) .

\begin{figure}
 \centering
\includegraphics[width=0.48\textwidth]{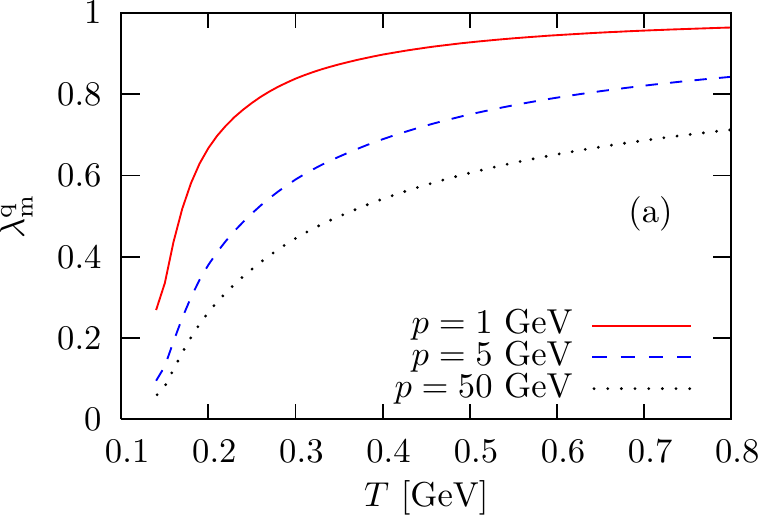}\\ 
\includegraphics[width=0.48\textwidth]{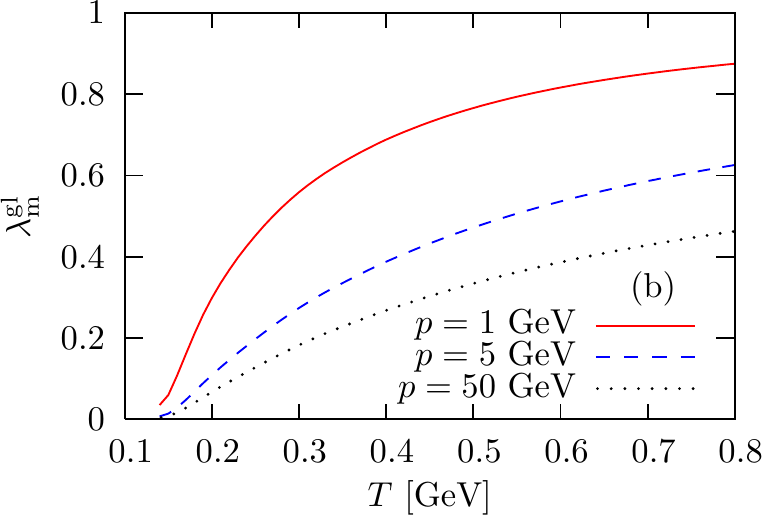}
 \caption{(Color online) The effective reduction $\lambda_{\rm m}$ of the scattering rates for different momenta of the charm quark as a function of the temperature for scattering off massive quasi-quarks (upper plot) and quasi-gluons (lower plot).}
 \label{fig:lambdath}
\end{figure}

\begin{figure}
 \centering
\includegraphics[width=0.48\textwidth]{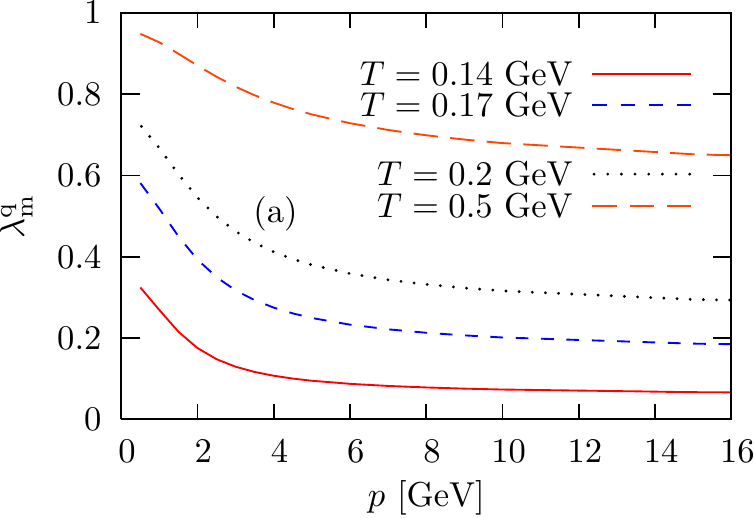}\\ 
\includegraphics[width=0.48\textwidth]{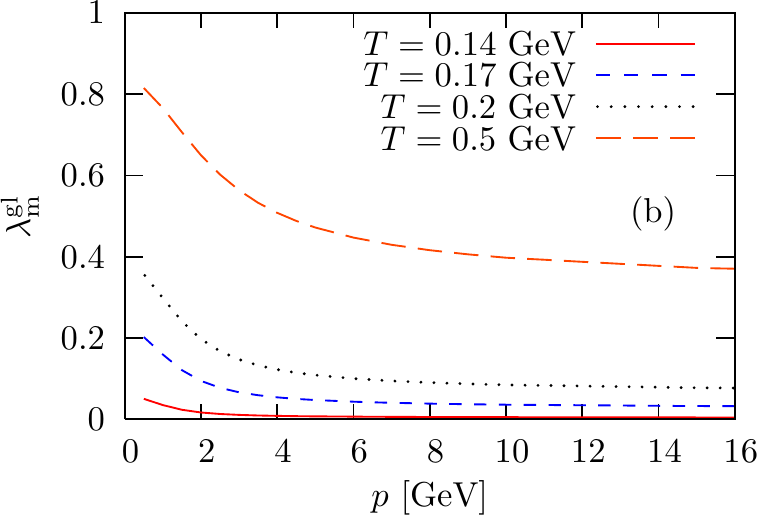}
 \caption{(Color online) The effective reduction $\lambda_{\rm m}$ of the scattering rates for different temperatures of the medium as a function of the momentum of the charm quark for scattering off massive quasi-quarks (upper plot) and quasi-gluons (lower plot). }
 \label{fig:lambdathpT}
\end{figure}

According to the calculation presented in appendix \ref{app:A} we define an effective reduction $\lambda_{\rm m}$ of the scattering rates via the ratio of the drag coefficients $A$ (see below for a definition of this quantity) for the case of massive quasiparticles as compared to massless light partons
\begin{equation}
 \lambda_{\rm m}(T,p)=\frac{A(m(T),p)}{A(m(T)=0,p)}\, ,
\end{equation}
with the quasiparticle masses as shown in Fig. \ref{fig:thmass}. We assume the same form of reduction for both types of scatterings, that of the charm quark with a massive quasi-quark, $Qq\to Qq$, and that with a massive quasi-gluon, $Qg\to Qg$  and in all channels, but use the respective thermal masses for quarks and gluons.
The thus obtained effective reductions of the scattering rates are shown for different momenta of the charm quark as a function of the temperature in Fig. \ref{fig:lambdath} and for different temperatures as a function of the momentum of the charm quark in Fig. \ref{fig:lambdathpT}.

The strong reduction of $\lambda_{\rm m}$ at low $T$ is due to the strong increase of the masses. Deep inside the high temperature QGP phase low-momentum heavy quarks are only slightly affected by the mass of their scattering partners. The scattering rates of high-momentum heavy-quarks, however, are significantly reduced even at high temperatures. Due to the assumption $m_{\rm gl}(T)=\sqrt{3}m_{\rm q}(T)$ the scattering of a heavy quark with a massive gluon is more strongly suppressed  than the scattering of a heavy quark with a massive light quark.

Contrary to the EPOS reduction of effective degrees of freedom the interpretation of the equation of state via thermal masses of quasiparticles leads to a momentum-dependent reduction of the scattering rates, see figure \ref{fig:lambdathpT}. One clearly observes a rapid decrease of $\lambda^{\rm q,gl}_{\rm m}$ with the momentum $p$ of the heavy quark for $p<4$~GeV, while $\lambda^{\rm q,gl}_{\rm m}$ is almost independent of the momentum for higher momenta, which comes from a saturation of the drag force at large momentum in the case of a running coupling constant. We can thus expect that for massive quasi-quarks the thermalization of low-momentum heavy-quarks is less suppressed by  scattering with partons of a  finite thermal mass than the energy loss at higher momenta. This should result in an elliptic flow, which (after fixing the $R_{\rm AA}$ at high momenta), is enhanced over a scenario where the effective reduction is only temperature-dependent but momentum-independent.

\section{Drag coefficient}\label{sec:drag}

\begin{figure}
 \centering
 \includegraphics[width=0.48\textwidth]{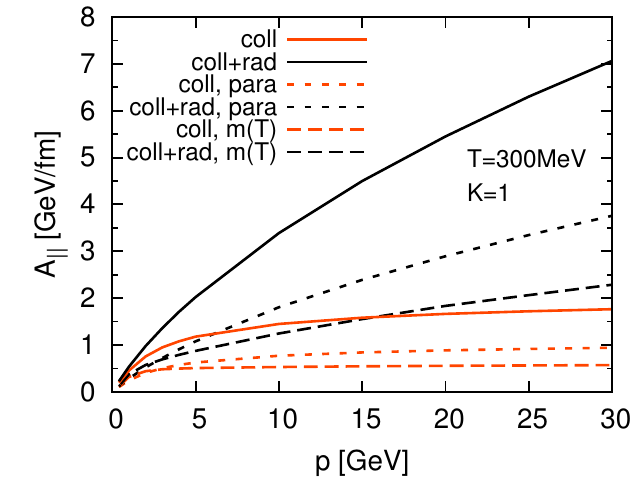}\\ 
\includegraphics[width=0.48\textwidth]{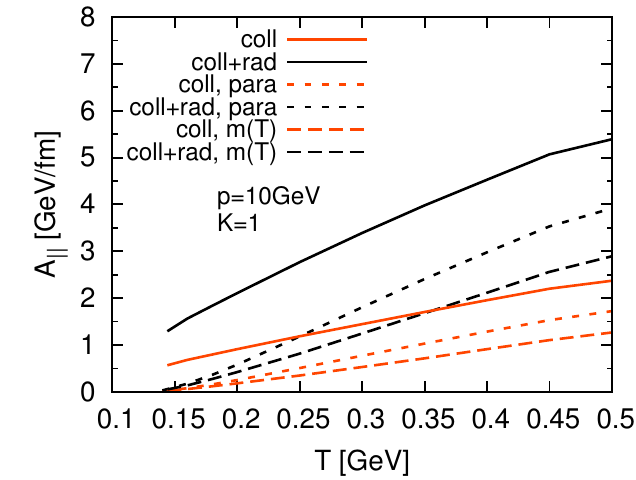}
 \caption{(Color online) The drag force for the three different representations of the QGP constituents: massless partons (solid), EPOS parametrization of the equation of state (short dashed) and massive quasiparticles (long dashed), as a function of the momentum (upper plot) and as a function of the medium temperature (lower plot). Results for the purely collisional energy loss (black) and for the collisional+radiative(LPM) are shown.}
 \label{fig:AdragnoK}
\end{figure}

\begin{figure}
 \centering
\includegraphics[width=0.48\textwidth]{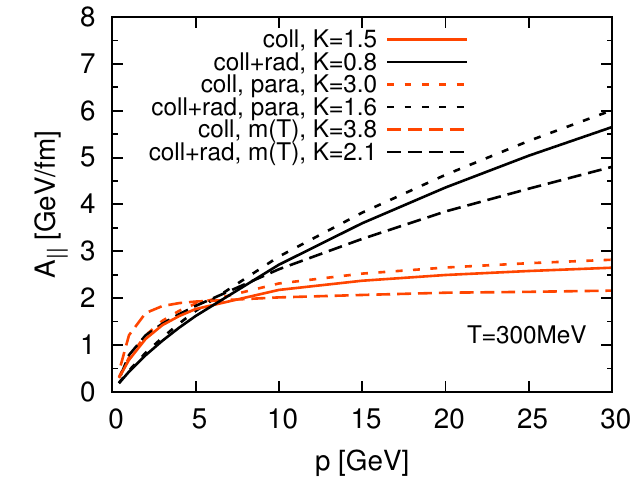}\\ 
\includegraphics[width=0.48\textwidth]{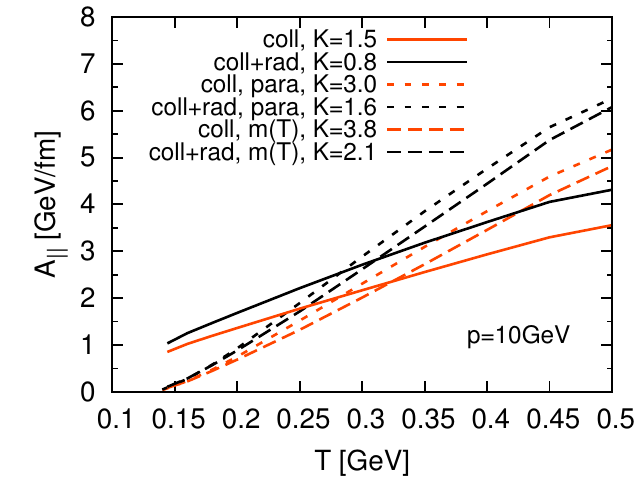}
 \caption{(Color online) Same as in Fig. \ref{fig:AdragnoK}, but the different scenarios are rescaled with the $K$-factor determined from an optimal description of the central $R_{\rm AA}$ data at LHC. See text for details.}
 \label{fig:AdragK}
\end{figure}

The drag coefficient is a relevant indicator for the energy loss which a heavy quark suffers in the medium. In Langevin/Fokker-Planck approaches it is the transport coefficient, which describes the average longitudinal momentum loss per time (isotropic medium), and thus is related to the drag force in Eq. (\ref{eq:x}) via
\begin{equation}
 A_{||}(\vec{p},T)=p_{||} A(p,T)=-\frac{{\rm d}p_{||}}{{\rm d}t}\bigg|_{T}\, .
\end{equation}

In Figs. \ref{fig:AdragnoK} and \ref{fig:AdragK} we present the drag force for three different assumptions about the degrees of freedom of the QGP constituents: first, the standard massless partons (solid lines), second, the EPOS parametrization of the equation of state (short dashed lines) and third, the description of the medium with massive quasiparticles (long dashed lines). In each of these plots the two energy loss mechanisms are shown, purely collisional (light, orange lines) and the collisional+radiative(LPM) scenario (black lines). Fig. \ref{fig:AdragnoK} shows the drag force at fixed temperature $T=300$~MeV as a function of the momentum of the charm quark (upper plot) and as a function of temperature at fixed momentum (lower plot). The same scenarios are shown in Fig. \ref{fig:AdragK}, but now including the global rescaling factors $K$, which are obtained from an optimal description of the $R_{\rm AA}$ around $p_T\sim 10$~GeV at $\sqrt{s_{\rm NN}}=2.76$~TeV Pb+Pb collisions at the LHC
(see section IV).

We observe in both figures that the collisional+radiative(LPM) energy loss increases with the momentum at high momenta, while the purely collisional energy loss mechanism leads to at most a logarithmic increase. The reduction of the energy loss in both scenarios in which the QGP constituents are represented consistently with the equation of state as compared to the standard case of massless partons is clearly visible in Fig. \ref{fig:AdragnoK}. The drag force for each energy loss mechanism is smallest for the medium consisting of massive quasiparticles. The temperature dependence of the reduction factor $\lambda$ becomes obvious in the lower plot of these Figs. \ref{fig:AdragnoK} and \ref{fig:AdragK}. While the drag force is finite at temperatures close to $T_c$ in the ideal gas of massless partons, it (almost) vanishes for the two other more realistic representations of the QGP constituents. In the PHSD model the temperature dependence of the strong coupling constant and its strong increase in the vicinity of $T_c$ counter balance the reduction of the drag force due to the large masses \cite{phsdHF}.

When one looks at the rescaled drag force in Fig. \ref{fig:AdragK} one sees immediately that the curves as function of momentum are closer together for each of the energy loss mechanisms than in Fig. \ref{fig:AdragnoK}. The determination of the $K$-factors in comparison to experimental data for the $R_{\rm AA}$ includes an integration over the entire space-time evolution and an averaging over several initial momenta. Despite the therefore very similar values of the $R_{\rm AA}$ (see section IV), the temperature dependence, see lower plot of Fig. \ref{fig:AdragK}, is still very different.
One observes in particular a stronger temperature-dependence for both energy loss scenarios inspired by the equation of state. In all of the rescaled scenarios the drag force  for the purely collisional scenario exceeds that for a collisional+radiative(LPM) interaction mechanism at low momenta $p\lesssim 5$~GeV. Due to the momentum dependence of the reduction factor $\lambda$ the enhancement of the low-momentum drag force for collisional energy loss is most pronounced in the case of massive quasiparticles.

We now proceed by applying our model to the full medium evolution in Pb+Pb collisions at $\sqrt{s}=2.76$~TeV and Au+Au collisions at $\sqrt{s}=200$~GeV.

\section{$R_{\rm AA}$ and $v_2$ at LHC and RHIC}\label{sec:results}

For the fully coupled evolution of heavy quarks and the QGP medium, we initialize the charm quarks at the nucleon-nucleon scattering points from the EPOS2 initial conditions. The initial transverse momentum spectra is obtained from FONLL calculations \cite{FONLL}. Nuclear shadowing has been included according to the EPS09 parametrization of the nuclear parton distribution functions \cite{Eskola:2009uj}. For the charm quark propagation in the medium we solve the same Boltzmann equation as in previous works, where the medium constituents are massless partons, but reduce the scattering rate according to the effective reduction factors $\lambda$ as discussed in the previous section. After the evolution in the medium the charm quarks hadronize at a given temperature via fragmentation predominantly at high $p_T$ and coalescence predominantly at low $p_T$.

\begin{figure}[tb]
\centering
\includegraphics[width=0.48\textwidth]{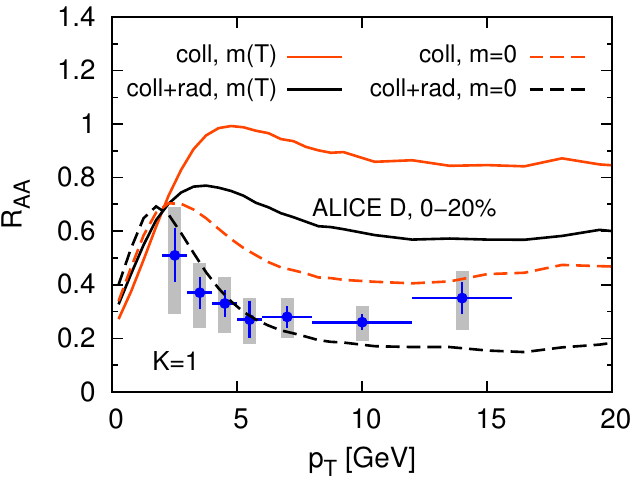}
\caption{(Color online) Comparison of the $D$ meson $R_{\rm AA}$ for a QGP consisting of massive quasiparticles (solid lines) and massless partons (dashed lines). The cross sections are not rescaled ($K=1$). Purely collisional (orange, light) and collisional+radiative(LPM) (black) energy loss scenarios are shown.}
 \label{fig:K1}
\end{figure}

In Fig. \ref{fig:K1} we show the $R_{\rm AA}$ of $D$ mesons for $\sqrt{s_{\rm NN}}=2.76$~TeV central Pb+Pb collisions in the case of massive quasiparticles in the QGP compared to the case where the QGP consists of massless partons. The decoupling temperature is $T=168$~MeV. In these calculations the cross sections are not rescaled, which corresponds to $K=1$. The case where the medium constituents are massless partons is shown with the dashed lines. Here, the purely collisional energy loss is evidently not strong enough to explain the experimental data. Including radiative corrections we find a good description for the intermediate $p_T$ range in the case of the massless QGP but the overall suppression at high $p_T$ is too large. These observations motivate the choice of $K$-factors. As expected, the strongly reduced drag force of charm quarks in a medium with massive quasiparticles leads to a substantial reduction of the energy loss of charm quarks for both energy loss scenarios, purely collisional and collisional+radiative(LPM). 

In the following, we use the $K$-factors determined by the $R_{\rm AA}$-data around $p_T\sim 10$~GeV at a decoupling temperature of $T_f=134$~MeV. Then this same $K$-factor is also applied for the case where the heavy-quark propagation 
hadronizes at $T=168$~MeV, which is the EPOS2 particlization temperature. By comparing results for the higher and lower decoupling temperatures one can see the effect of the late-stage evolution, during which the according degrees of freedom are still in accordance with the interpretation of the equation of state. 

\begin{figure}[tb]
\centering
\includegraphics[width=0.48\textwidth]{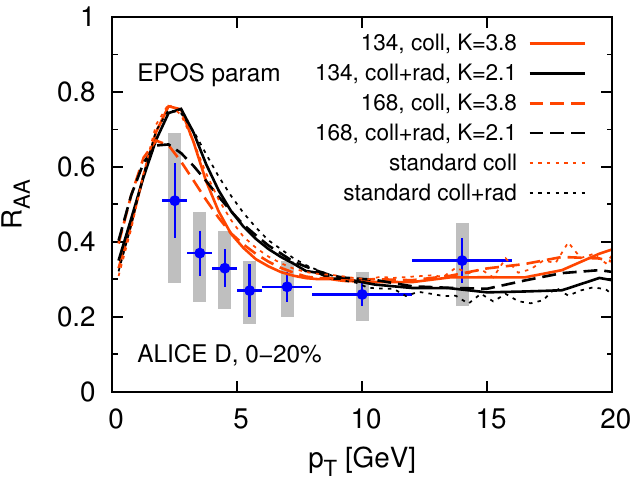}\\
\includegraphics[width=0.48\textwidth]{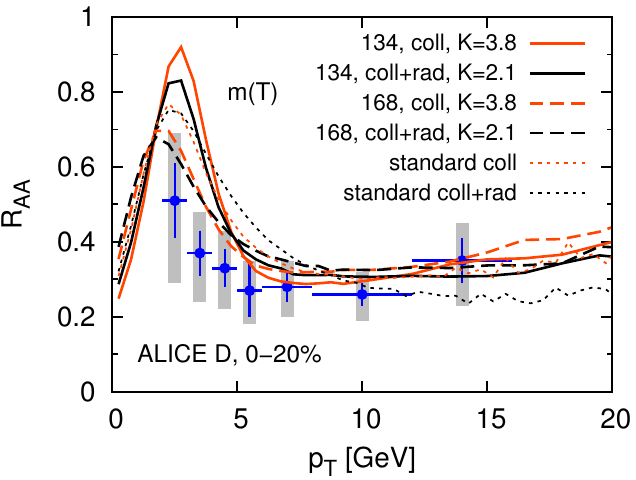}
\caption{(Color online) $D$ meson $R_{\rm AA}$ in central $\sqrt{s_{\rm NN}}=2.76$~TeV Pb+Pb collisions. In the upper plot the EPOS parametrization of the equation of state is used, in the lower plot the scenario with massive quasiparticles. Shown are curves for purely collisional and collisional+radiative(LPM) energy loss and for the two decoupling-temperatures $T=134$~MeV and $T=168$~MeV. Standard curves refer to the case of massless light partons.}
 \label{fig:LHCRAA}
\end{figure}

\begin{figure}[tb]
\centering
\includegraphics[width=0.48\textwidth]{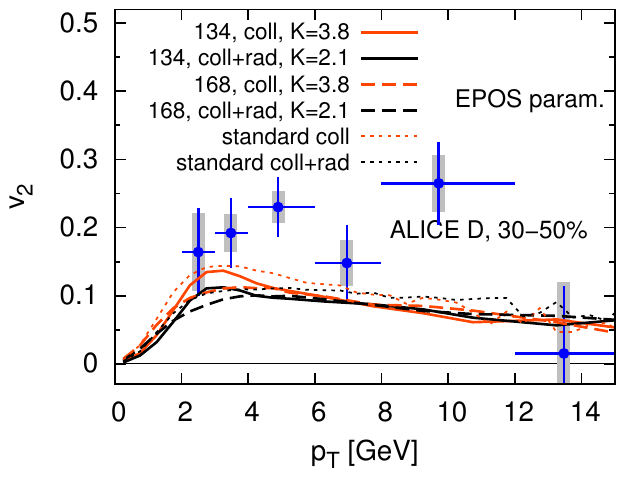}\\
\includegraphics[width=0.48\textwidth]{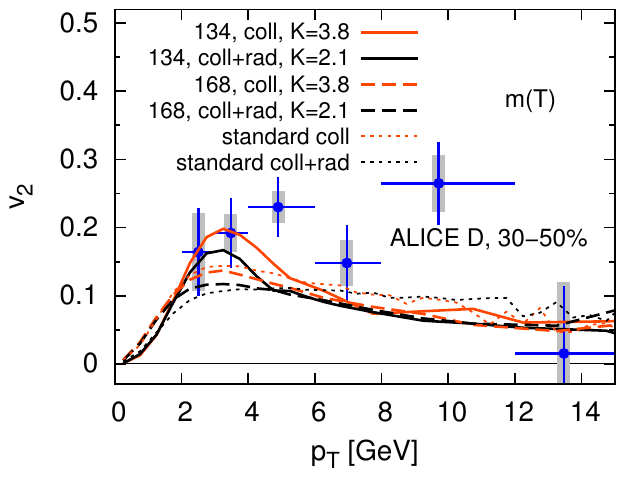}
\caption{(Color online) $D$ meson $v_2$ in $30-50$\% most central $\sqrt{s_{\rm NN}}=2.76$~TeV Pb+Pb collisions. In the upper plot the EPOS parametrization of the equation of state is used, in the lower plot the scenario with massive quasiparticles. Shown are curves for purely collisional and collisional+radiative(LPM) energy loss and for the two decoupling temperatures $T=134$~MeV and $T=168$~MeV. Standard curves refer to the case of massless light partons.}
 \label{fig:LHCv2}
\end{figure}

In Fig.~\ref{fig:LHCRAA} we show the $R_{\rm AA}$ of $D$ mesons in central collisions at LHC for both interpretations of the equation of state, the EPOS parametrization (upper plot) and the massive quasiparticle interpretation of the light partons (lower plot) as well as for the standard scenario (massless quarks with $T_c = 155$~ MeV and $K_c=1.22$, $K_{c+r}=0.6$\footnote{We note, that these $K$ factors differ from the ones in previous publications, which were chosen such that intermediate and high-$p_T$ $R_{\rm AA}$ data were optimally reproduced, whereas here we focus on reproducing $R_{\rm AA}(p_T\sim10\,{\rm GeV})\sim 0.3$.}). It is seen that $K$ factors can be determined such that the experimental data above $p_T\gtrsim8$~GeV can equally well be described by both representations of the medium constituents. At high momenta, the decoupling temperatures do not affect the results. Below $p_T\lesssim5$~GeV differences become more prominent. In both cases, EPOS parametrization of the equation of state or massive quasiparticles, the peak around $p_T\sim1.5-2.5$~GeV is higher and shifted toward larger $p_T$ if the evolution is stopped at a later temperature, since the radial flow in the medium increases with the evolution time. 
Since the effective reduction of the scattering rates in the scenario with massive quasiparticles depends on the incoming momentum of the charm quark and scatterings of low-momentum charm quarks are less suppressed the coupling to the radial flow is stronger in this case. This is especially true for a prolonged evolution in the low temperature phase. Toward higher $p_T$ we can see a slight upward trend of the $R_{\rm AA}$ for the purely collisional energy loss scenario. 

Figs.~\ref{fig:LHCv2} shows the elliptic flow $v_2$ of $D$ mesons in the $30-50$~\% most central collisions at LHC. It can be seen again that due to the momentum dependence of the effective reduction, $\lambda_{\rm m}^{\rm g,q}$, in a massive quasiparticle picture the coupling to the flow of the medium is stronger at low momenta than it is the case for the EPOS parametrization of the equation of state. The evolution at lower temperatures can again enhance the $v_2$ compared to a higher decoupling temperature. 
\begin{figure}[tb]
\centering
\includegraphics[width=0.48\textwidth]{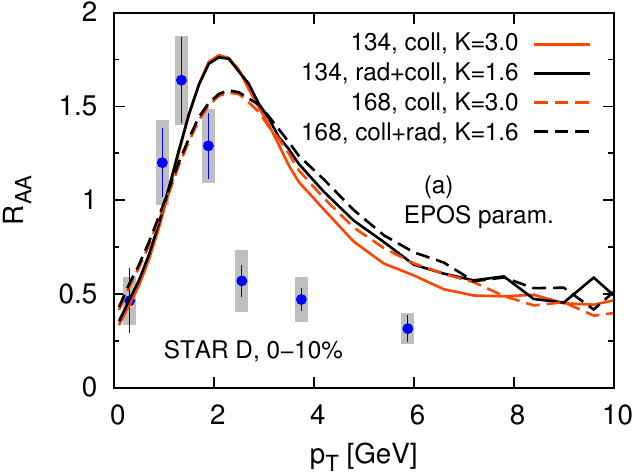}\\
\includegraphics[width=0.48\textwidth]{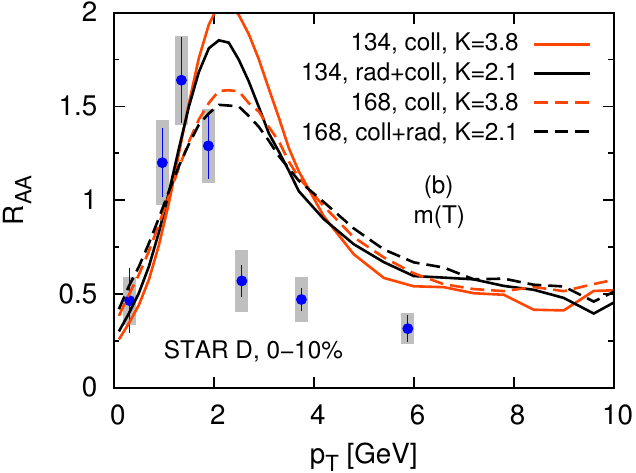}
\caption{(Color online) $D$ meson $R_{\rm AA}$ in central $\sqrt{s_{\rm NN}}=200$~GeV $Au+Au$ collisions. In the upper plot the EPOS parametrization of the equation of state is used, in the lower plot the scenario with massive quasiparticles. Shown are curves for purely collisional and collisional+radiative(LPM) energy loss and for the two decoupling-temperatures $T=134$~MeV and $T=168$~MeV.}
 \label{fig:RHICRAA}
\end{figure}
\begin{figure}[tb]
\centering
  \includegraphics[width=0.48\textwidth]{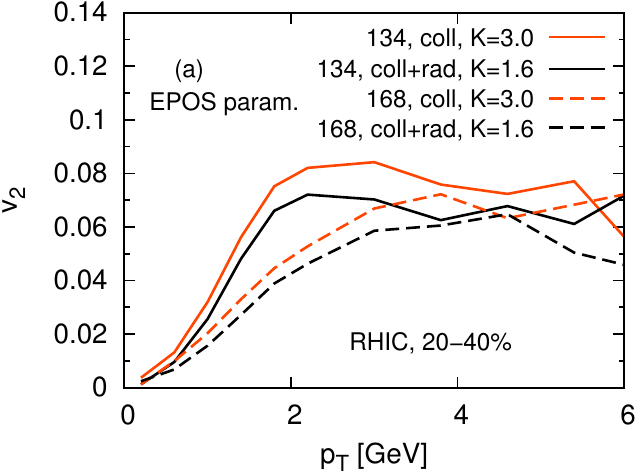}\\
\includegraphics[width=0.48\textwidth]{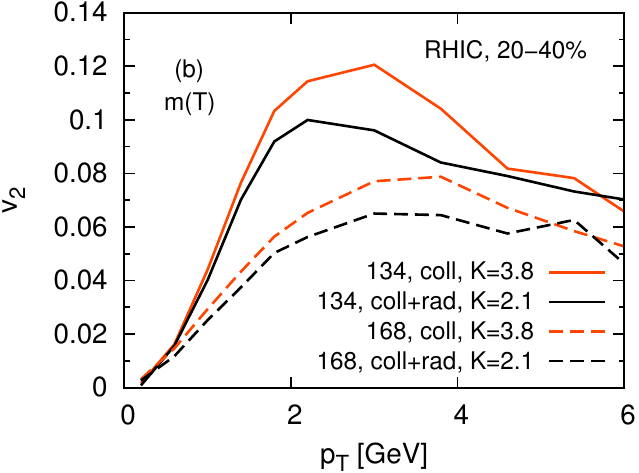}
\caption{(Color online) $D$ meson $v_2$ in $20-40$\% most central $\sqrt{s_{\rm NN}}=200$~GeV $Au+Au$ collisions. In the upper plot the EPOS parametrization of the equation of state is used, in the lower plot the scenario with massive quasiparticles. Shown are curves for purely collisional and collisional+radiative(LPM) energy loss and for the two decoupling-temperatures $T=134$~MeV and $T=168$~MeV.}
 \label{fig:RHICv2}
\end{figure}
For completeness, we show our results for $R_{\rm AA}$ and $v_2$ at top-RHIC energy in Figs. \ref{fig:RHICRAA} and \ref{fig:RHICv2}. We observe the same trends with respect to the representation of the medium constituents, the decoupling temperature and the energy loss mechanism as for the LHC. In order to determine the $K$-factors at RHIC, data at higher $p_T$ for the $D$ meson $R_{\rm AA}$ would be very helpful. The now available data covers a $p_T$ range which has not been included in the $K$-factor determination at LHC. Here, we take therefore the same $K$-factors as determined at the LHC and investigate the consequences at the lower energy although there are some indications that slightly higher $K$-factors improve the comparison to the data at RHIC \cite{Nahrgang:2014ila,Nahrgang:2014vza}. 
It seems, however, that no model can be tuned in such a way that it is able to cope with the data, which might be a sign for an excess of radial
flow in the background medium.

\section{Conclusions}\label{sec:conclusions}
In the present work, we have investigated the possibility to  couple consistently the heavy-flavor dynamics to the fluid dynamical evolution of the light bulk particles in ultrarelativistic heavy-ion collisions. While today in most models the equation of state is taken realistically as the lattice QCD equation of state, thermal scattering partners are mostly sampled from the distribution of a noninteracting, ideal gas of relativistic, massless partons. This characterization of the nature of the quasiparticles might only be justified at extremely high temperatures where the Stefan-Boltzmann limit will eventually be reached. Here we looked into two different representations of the degrees of freedom in the QGP, which both reproduce the correct lattice QCD equation of state. One is a parametrization of the lattice equation of state in terms of partonic and hadronic degrees of freedom and is used  in the fluid dynamical evolution of the EPOS2 approach. The other is the description of the medium constituents as massive quasiparticles. The thermal masses have been obtained by fitting the lattice equation of state. They show a strong increase at lower temperatures. In the first case, the scattering rate of the heavy quarks with the medium constituents is reduced by the fraction of hadronic color-neutral degrees of freedom in the medium, which is a function of the medium temperature. In the second case, we derived an effective reduction of the scattering rate via a comparison of the drag force for massive to that of massless medium constituents. Here the reduction factor depends on the temperature, the momentum of the heavy quark and whether the heavy quark scatters with a quark or a gluon.

We could show that the high momentum part of the $R_{\rm AA}$ remains unchanged after a rescaling of the rates with a global $K$-factor although the temperature dependence of the drag force is very different in a realistic description of the medium constituents compared to massless partons. Due to the reduced drag force, the $K$-factors needed to reproduce the $R_{\rm AA}$ data are found to be significantly larger than in previous calculations with massless partons. Differences are visible at lower transverse momenta in the $R_{\rm AA}$ and in the elliptic flow $v_2$, where the late stage evolution with low temperatures is more important and the scattering rate significantly reduced. Particularly interesting is the case of massive quasiparticles, where the additional momentum dependence of the reduction factor leads to a very pronounced impact of the late stage evolution. The coupling to the radial flow and the elliptic flow of the underlying medium is enhanced after a global rescaling of the scattering rates by a $K$-factor.

Since the central $R_{\rm AA}$ data at intermediate and high transverse momentum is currently used to calibrate the energy loss model, further observables like the elliptic flow studied here are sensitive to the  representation of the medium constituents, albeit the current precision of the data is not good enough to distinguish these effects. It would be interesting to look at azimuthal correlations and higher-order flow harmonics \cite{Nahrgang:2013saa,Nahrgang:2014vza}  in future work. For this, we will use the upgraded EPOS3 version which includes a coupled initial state for fluid dynamics and heavy-flavor production and a viscous fluid dynamical evolution.

The approach of massive quasiparticles seems one of the more realistic approaches to understand the thermodynamics, i.e. the equation of state of QCD, which can also be used as a foundation to study heavy-flavor dynamics. This approach should be followed in more detail to include the strong coupling effects leading to enhanced thermal masses of the light quasiparticles also with respect to the heavy-flavor interaction. The representation of the medium constituents by massless partons is an unrealistic approximation for temperatures which are reached in heavy-ion collisions.

 \section*{Acknowledgements}
M.N. was supported by a fellowship within the Postdoc-Program of the German Academic Exchange Service (DAAD). This work was supported by the U.S. department of Energy under grant DE-FG02-05ER41367. We are grateful for support from  the Hessian LOEWE initiative Helmholtz International 
Center for FAIR, ANR research program ``hadrons @ LHC''  (grant ANR-08-BLAN-0093-02),  TOGETHER project R\'egion Pays de la Loire and I3-HP. The authors appreciate fruitful discussions with Steffen Bass, Hamza Berrehrah, Elena Bratkovskaya, Marcus Bluhm, Krzysztof Redlich and Taesoo Song.

\appendix
\section{Heavy quark - massive light-quark scattering}\label{app:A}
We start from the definition of the drag force in \cite{Svetitsky:1987gq}
\begin{align}
A_i(\vec{p})=&\frac{1}{16 (2\pi)^5 E_p} \int \frac{d^3q}{E_q} f(\vec{q}) \int \frac{d^3q'}{E_{q'}} \int 
\frac{d^3p'}{E_{p'}} (p-p')^i \nonumber \\
&\times\delta^{(4)}(P_{\rm in}-P_{\rm fin}) \times \frac{\sum |{\cal M}|^2}
{\gamma_Q \gamma_p}\,,
\end{align}
where $\gamma_Q$ is the degeneracy of the heavy quark ($\gamma_Q=6$), 
$\gamma_p$ is the degeneracy of the light parton, $\vec{p}$($\vec{q}$) is the incoming
momentum of the heavy quark  (light parton), $\vec{p}\,'$ ($\vec{q}\, '$) is the final momentum.
 $E_x$ is the energy associated to the 
momentum $\vec{x}$. $m_Q$ is the mass of the heavy quark while $m_q$ is the mass of the light
parton. 
The drag force can easily be brought to a covariant form, 
as seen by solving the Fokker Planck equation
in the absence of diffusion ($B_{ij}=0$.
\begin{equation}
\frac{\partial f}{\partial t} = 
\frac{\partial}{\partial p_i} \left(A_i f\right)
\quad\Rightarrow\quad \frac{d\langle p_i\rangle_f}{dt} =-
\int d^3p\,\left(A_i f\right)\,
\end{equation}
where f is the particle distribution in momentum space.
In particular, if we take $f=\delta(\vec{p}-\vec{p}_0)$, we obtain
\begin{equation}
\frac{d p_{0,i}}{dt}=- A_i(\vec{p}_0)
\quad\Rightarrow \quad \frac{d \vec{p}_{0}}{d\tau} = -\frac{E_p}{m_Q} \vec{A}=-\vec{{\cal A}}\,,
\end{equation}
where $\tau$ is the eigentime of the heavy quark. $\frac{d \vec{p}_{0}}{d\tau}$ is the
spatial part of a covariant quantity. 
Adding the temporal part we can define 
the covariant ${\cal A}^\mu$
\begin{align}
{\cal A}^\mu(\vec{p})=&\frac{-1}{16 (2\pi)^5 m_Q} \int \frac{d^3q}{E_q} f(\vec{q}) 
\int \frac{d^3q'}{E_{q'}} \int 
\frac{d^3p'}{E_{p'}} (q-q')^\mu\nonumber \\
&\times \delta^{(4)}(P_{\rm in}-P_{\rm fin}) \times \frac{\sum |{\cal M}|^2}
{\gamma_Q \gamma_p}\,,
\end{align}
using $(p-p')^\mu=-(q-q')^\mu$.  Introducing
\begin{equation}
a^\mu:=\int \frac{d^3q'}{E_{q'}} \int 
\frac{d^3p'}{E_{p'}} (q-q')^\mu \delta^{(4)}(P_{\rm in}-P_{\rm fin}) \times 
\frac{\sum |{\cal M}|^2}
{\gamma_Q \gamma_p}\,,
\end{equation}
which is covariant as well we can evaluate  $a^\mu$ and 
\begin{equation}
{\cal A}^\mu= -\frac{1}{16 (2\pi)^5 m_Q} \int\frac{d^3q}{E_q} f(\vec{q}) a^\mu
\label{fiA6}
\end{equation}  
in different frames. 

\subsection{Evaluation of $a^\mu$}
We will evaluate $a^\mu$ in the heavy quark-light parton  c.m. frame (dubbed  $a_{cm}$) and
${\cal A}^\mu$ in the frame  where the heavy quark is at rest. $a_{cm}^0=0$ by construction and
\begin{equation}
\vec{a}_{cm}= \int \frac{d^3q'}{E_{q'}E_{p'}} \delta(\sqrt{s}- E_{q'}-E_{p'})
\left(\vec{q}-\vec{q}'\right) \frac{\sum |{\cal M}|^2}
{\gamma_Q \gamma_p}\,.
\label{amu}
\end{equation}
Using $E_{p'}=\sqrt{m_Q^2+p_{\rm rel}^2}$ and $E_{q'}=\sqrt{m_q^2+p_{\rm rel}^2}$
we obtain 
\begin{equation}
\delta(\sqrt{s}- E_{q'}-E_{p'})=\frac{\delta(q'-p_{\rm rel})}
{\frac{p_{\rm rel}}{E_{q'}}+\frac{p_{\rm rel}}{E_{p'}}}=\frac{E_{q'} E_{p'}}
{p_{\rm rel}\sqrt{s}}\times \delta(q'-p_{\rm rel})\,,
\end{equation}
and thus
\begin{equation}
\vec{a}_{cm}= \frac{p_{\rm rel}}{\sqrt{s}}\int d\Omega_{q'}  
\left(\vec{q}-\vec{q}'\right) \frac{\sum |{\cal M}|^2(s,t)}{\gamma_Q \gamma_p}\,.
\end{equation}
Due to symmetry, $\vec{a}_{cm}\parallel \hat{q}_{cm}$, where $\hat{q}$
is the unit vector in the direction of the light parton in the c.m., and we write
$$\vec{a}_{cm} =a_{cm} \hat{q}\,,$$
with
\begin{equation}
a_{cm}=\frac{p_{\rm rel}}{\sqrt{s}}\int d\Omega_{q'}  
\left(\vec{q}-\vec{q}'\right)\cdot \hat{q}  \frac{\sum |{\cal M}|^2(s,t)}{\gamma_Q \gamma_p}\,.
\end{equation}
Introducing the angle $\theta$ between $\vec{q}$ and $\vec{q}'$ leads to
\begin{equation}
a_{cm}=\frac{2\pi p_{\rm rel}^2(s)}{\sqrt{s}}
\int d\cos\theta  \left(1-\cos\theta\right) \frac{\sum |{\cal M}|^2(s,t)}{\gamma_Q \gamma_p}\,.
\end{equation}
We define 
\begin{equation}
m_1(s):=
\int_{-1}^{+1} d\cos\theta  \frac{1-\cos\theta}{2} 
\frac{\sum |{\cal M}|^2(s,t)}{\gamma_Q \gamma_p}\,,
\end{equation}
which is positive defined, implying a force along $-\hat{q}$, which appears natural if one goes in the rest frame of the heavy quark. We obtain
\begin{equation}
a_{cm}^\mu=
\left\{\begin{array}{ll}
0 & \text{for $\mu=0$}\\
\frac{4\pi p_{\rm rel}^2(s) m_1(s)}{\sqrt{s}} \times \hat{q}_{cm}^\mu
& \text{for $\mu\neq 0$}
\end{array}\right.\,.
\label{fiA13}
\end{equation}

\subsection{Calculation of $m_1$}
For the evaluation of $m_1$  we
introduce $v=\cos\theta-1$:
$$m_1(s):=-\int_{-2}^{0} \frac{v}{2} \frac{\sum |{\cal M}|^2(s,t(v))}{\gamma_Q \gamma_p}\,dv\,.$$
In the c.m., one finds $t=(p-p')^2=-2 p_{\rm rel}^2 (1-\cos\theta)=2 p_{\rm rel}^2 v$, which yields
\begin{equation}
m_1(s):=\frac{1}{8 p_{\rm rel}^4}\int_{-4p_{\rm rel}^2}^{0}  
\frac{\sum |{\cal M}|^2(s,t)}{\gamma_Q \gamma_p}\, (-t) dt\,.
\label{eq_def_m1}
\end{equation}

The scattering amplitude ${\cal M}$  between a heavy and a light quark reads (conventions of Itzikson and Zuber) 
\begin{align}
{\cal M}=& g^2 \sum_\lambda t_{a,a'}^\lambda t_{b,b'}^\lambda  \\ &\times  \frac{\bar{u}^{(s_{p'})}(p') \gamma^\mu u^{(s_{p})}(p)\,
g_{\mu\nu}\bar{u}^{(s_{q'})}(q') \gamma^\nu u^{(s_q)}(q)}{t-m_g^2}\nonumber \,,
\end{align}
where $a$ ($b$) is the initial color of the heavy quark (light quark), and $a'$ ($b'$) are the final colors. Up to the color factor
the matrix element is identical to that for $e^-\mu^-$ scattering and can be found in standard text books.
\begin{equation*}
\sum |{\cal M}|^2= 8 g^4\times {\rm col} \times \frac{s_-^2+u_-^2+2(m_Q^2+m_q^2) t}{(t-m_g^2)^2} \nonumber \,.
\end{equation*}
with $s_-:=s-m_Q^2-m_q^2$ and $u_-:=u-m_Q^2-m_q^2$. 
The color factor ``col'' is evaluated to
\begin{align}
 {\rm col}:=&\sum_{a,a',b,b'} \left|\sum_{\lambda=1}^{8} t_{a a'}^\lambda t_{b b'}^\lambda\right|^2=
\sum_{\lambda,\lambda'=1}^{8}\left[{\rm Tr}(t^\lambda t^{\lambda'})\right]^2\nonumber \\ =&
\sum_{\lambda,\lambda'=1}^{8} \frac{\delta_{\lambda,\lambda'}}{4}=2\,.
\end{align}
The matrix element reads therefore
\begin{align}
\frac{\sum |{\cal M}|^2}{\gamma_q \gamma_Q}=&\frac{16 g^4}{\gamma_q \gamma_Q}\nonumber \times 
\frac{s_-^2+u_-^2+2(m_Q^2+m_q^2) t}{(t-m_g^2)^2}
\nonumber\\
=& \frac{256 \pi^2 \alpha_s^2}{\gamma_q \gamma_Q} \times 
\frac{s_-^2+u_-^2+2(m_Q^2+m_q^2) t}{(t-m_g^2)^2}\nonumber
\label{trans_matrix_Qq}
\end{align}
with $\gamma_Q=\gamma_q=6$ and we obtain
\begin{equation}
m_1(s)=\frac{8 \pi^2}{9 p_{\rm rel}^4}\int_{-4p_{\rm rel}^2}^{0} \!\!\!\! (-t) dt \, \alpha_s^2\,\frac{s_-^2+u_-^2+2(m_Q^2+m_q^2) t}{(t-m_g^2)^2}\,,
\end{equation}
where $\alpha_s$ is $t$-dependent.
\begin{figure}[tb]
\includegraphics[width=0.48\textwidth]{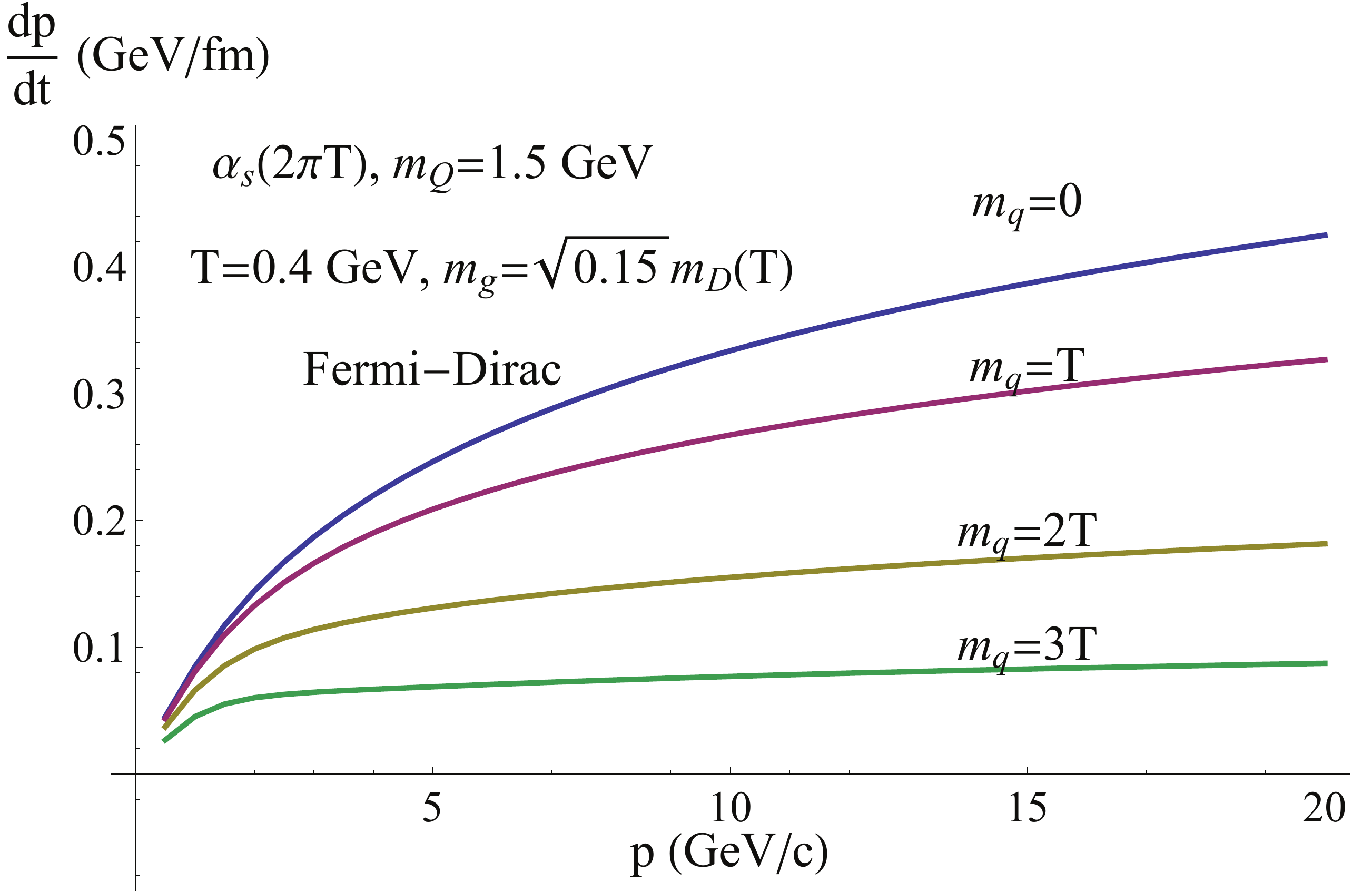}\\
\includegraphics[width=0.48\textwidth]{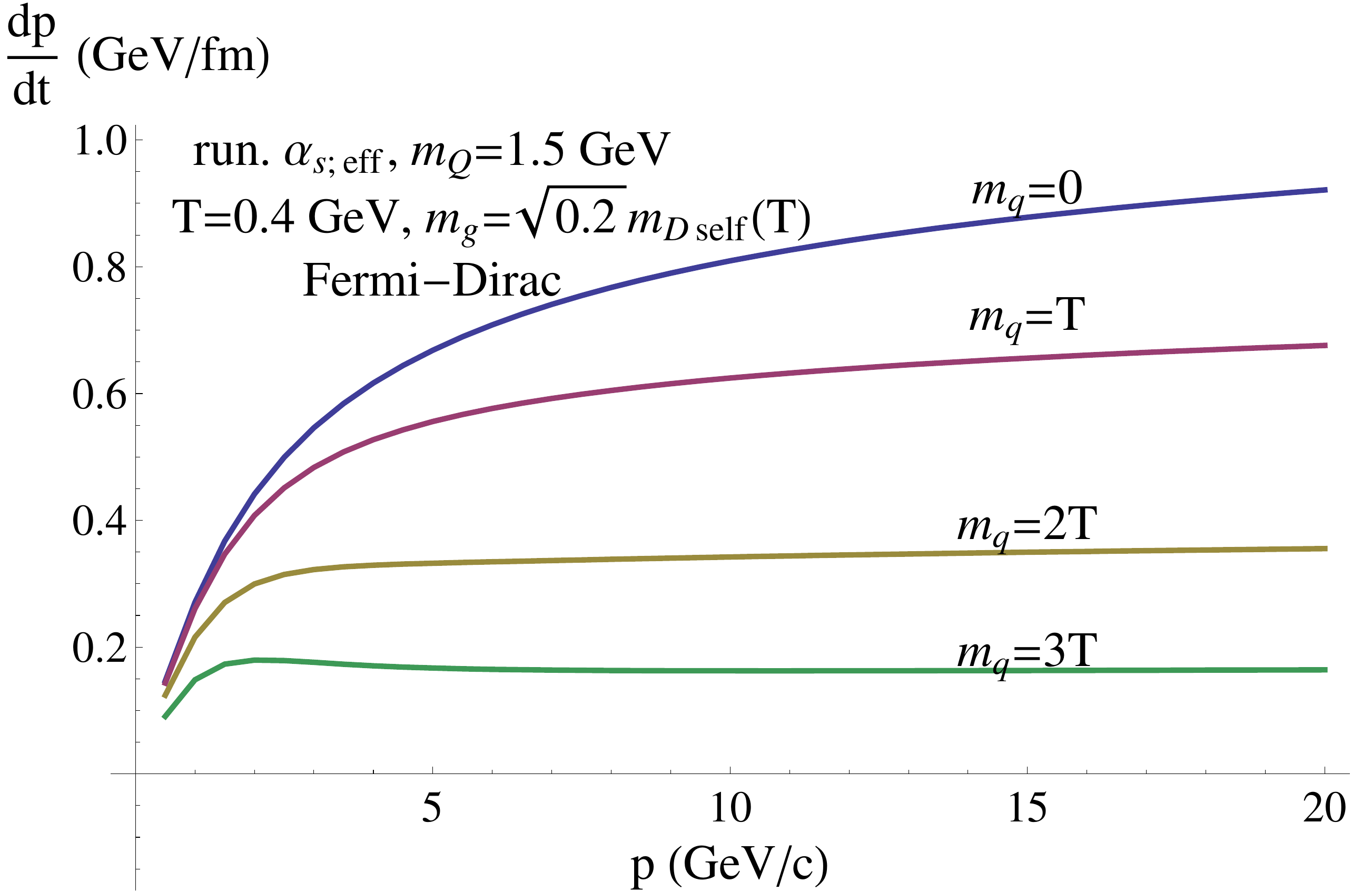}
\caption{Top: contribution of quarks to the drag of $c$-quarks as a function of $p$ with
parameters corresponding to model C of~\cite{Gossiaux:2008jv}; bottom: same for model E}
\label{fig_drag_C_et_E}
\end{figure}

\subsection{${\cal A}^\mu$ in the heavy quark rest system}
In the c.m. system $a^\mu_{cm}$ is given by Eq.\ref{fiA13}.
The boost matrix $\Lambda$ from the c.m. to the rest frame of the heavy quark is given by
$$
\Lambda(\vec{u}) = 
\left(\begin{array}{cc}
u^0 & -\langle\vec{u} \\
-\vec{u}\rangle & 1 + \frac{\vec{u}\rangle \langle\vec{u}}{1+u^0} 
\end{array}\right)\,,$$
where $(u^0,\vec{u})$ is the 4-velocity of the rest frame of the $c$-quark seen from the c.m. frame.
$\vec{u}$ is the opposite of the 4-velocity of the c.m. in the $c$-rest frame, i.e.
$$\vec{u}=-\frac{\vec{q}_r}{\sqrt{s}}\quad\text{and}\quad u^0=\frac{E_{q_r}+m_Q}{\sqrt{s}}\,,$$
with $s=(m_Q+E_{q_r})^2-q_r^2=m_Q^2+m_q^2+2 m_Q E_{q_r}$, where the subscript $r$ indicates 
that the quantities are considered in the rest frame of the $c$-quark. We obtain
$$
\Lambda=\frac{1}{\sqrt{s}}
\left(\begin{array}{cc}
E_{q_r}+m_Q & \langle\vec{q}_r \\
\vec{q}_r\rangle & \sqrt{s} + \frac{\vec{q}_r\rangle \langle\vec{q}_r}
{\sqrt{s}+E_{q_r}+m_Q} \end{array}\right)
$$
and
\begin{align}
a_r=&\frac{4\pi p_{\rm rel}^2(s) m_1(s)}{s} \nonumber \\ \times &
\left(\begin{array}{cc}
E_{q_r}+m_Q & \langle\vec{q}_r \\
\vec{q}_r\rangle & \sqrt{s} + \frac{\vec{q}_r\rangle \langle\vec{q}_r}
{\sqrt{s}+E_{q_r}+m_Q} \end{array}\right)\cdot\left(\begin{array}{c}
0\\ \hat{q}_{cm}\end{array}\right)\nonumber \\
=&\frac{4\pi p_{\rm rel}^2(s) m_1(s)}{s}
\left(\begin{array}{c}
q_r \\ \left(E_{q_r}+m_Q\right)\hat{q}_r \end{array}\right)\,
\end{align}
where we have used that  $\hat{q}_{cm}=\hat{q}_{r}$, i.e. the direction of $\vec{q}_r$ is
not affected when going to the c.m. frame. In the heavy quark rest system ${\cal A}^\mu$, defined in eq. \ref{fiA6},
is given by 
\begin{align}
{\cal A}^\mu_{\rm rest}
=&-\frac{1}{8(2\pi)^4m_Q}
\int \frac{p_{\rm rel}^2(s) m_1(s)}{s}\nonumber \\ \times& \left[\int d\Omega_{\vec{q}}
\left(q \delta^{\mu 0}+(E_{q}+m_Q) \hat{q}_i \delta^{\mu i}\right) f_r(\vec{q})\right]
\frac{q^2 dq}{E_{q}}\nonumber \\
=&-\frac{1}{4(2\pi)^3m_Q}\int 
\frac{p_{\rm rel}^2(s) m_1(s)}{s}\nonumber \\\times& \left[q f_0(q) \delta^{\mu 0} + 
(E_{q}+m_Q) f_1(q) \delta^{\mu i} \left(\hat{u}\right)_i\right]\frac{q^2 dq}{E_{q}}
\nonumber\\
=&-\left(A_{\rm rest}^0 + A_{\rm rest}^v \hat{u}\right)^\mu\,,
\end{align}
where we have introduced
\begin{align}
4\pi f_0(q)=&\int d\Omega_{\vec{q}} f\left(\frac{u^0 E_q- u q \cos\theta}{T}\right) \nonumber \\ =&
2\pi \int d\cos\theta f\left(\frac{u^0 E_q- u q \cos\theta}{T}\right)  \, 
\end{align}
and
\begin{align}
 4\pi f_1(q)=&\int d\Omega_{\vec{q}} f\left(\frac{u^0 E_q- u q \cos\theta}{T}\right)\cos\theta  \\ =&
2\pi \int d\cos\theta f\left(\frac{u^0 E_q- u q \cos\theta}{T}\right)\cos\theta \nonumber  \, .
\end{align}
In the latter equation we used the fact that for the $\mu=i$ components symmetry requires that ${\cal A}^i_{\rm rest}$ should be directed along $\vec{u}$. Hence $\vec{\cal A}_{\rm rest}=\left(\vec{\cal A}_{\rm rest}\cdot \hat{u}\right)\,
\hat{u}$.

For a Juettner-Boltzmann distribution $f_0$ and $f_1$ are given by
\begin{align}
f(\vec{q})=e^{-\frac{u^0 E_q}{T}+\frac{\vec{q}\cdot\vec{u}}{T}} \Rightarrow
f_0(q)=&\frac{e^{-\frac{u^0 E_q}{T}}}{2} \int_{-1}^{+1} d\cos\theta
e^{\frac{q u \cos\theta}{T}} \nonumber \\ =&
e^{-\frac{u^0 E_q}{T}} \times \frac{\sinh \alpha}{\alpha}\,,
\end{align}
and
\begin{equation}
f_1(q)=e^{-\frac{u^0 E_q}{T}} \times \frac{\partial}{\partial \alpha}\frac{\sinh \alpha}{\alpha}\,.
\end{equation}
with $\alpha:=\frac{q u}{T}$.

For the Fermi-Dirac distribution, which we use for our calculation,  there is also an analytical solution  for the moments $f_0$ and $f_1$.
\begin{equation}
f(\vec{q})=\frac{1}{e^{\frac{u^0 E_q}{T}-\frac{\vec{q}\cdot\vec{u}}{T}}+1}
\quad \Rightarrow\quad
f_0(q)=\frac{1}{2}\int_{-1}^{+1} \frac{dv}{e^{\frac{u^0 E_q-u q v}{T}}+1}\,,
\end{equation}
where $v=\cos\theta$. 
We rewrite this equation, introducing $\alpha=\frac{u q}{T}$ and $\beta=\frac{u^0 E_q}{T}$.
\begin{align}
f_0(q)=&\frac{1}{2}\int_{-1}^{+1} \frac{e^{\frac{u q v}{T}}\,dv}{e^{\frac{u^0 E_q}{T}}+e^{\frac{u q v}{T}}}=
\frac{1}{2}\int_{-1}^{+1} \frac{e^{\alpha v}\,dv}{e^{\beta}+e^{\alpha v}}
\nonumber\\
=&\frac{1}{2\alpha} \int_{e^{-\alpha}}^{e^{+\alpha}} \frac{d\left(e^{\alpha v}\right)}{e^{\beta}+e^{\alpha v}}
=\frac{1}{2\alpha}\ln\left(\frac{e^{\beta}+e^{\alpha}}{e^{\beta}+e^{-\alpha}}\right)\,.
\end{align}
For $f_1$, we get
\begin{equation}
f_1(q)=
\frac{1}{2}\int_{-1}^{+1} \frac{v \frac{e^{\alpha v}}{e^\beta}\,dv}{1+\frac{e^{\alpha v}}{e^\beta}}=
\frac{1}{2}\sum_{n=1}^{+\infty} (-1)^{n+1} \int_{-1}^{+1} v \left(\frac{e^{\alpha v}}{e^\beta}\right)^n\,dv
\end{equation}
Using the variable change $\tilde{v}=n \alpha v$, one gets
\begin{align}
f_1(q) =& \frac{1}{2} \sum_{n=1}^{+\infty} \frac{(-1)^{n}}{(n \alpha)^2 e^{n \beta}}  
\left[(1-\tilde{v}) e^{\tilde{v}}\right]_{-n \alpha}^{+ n\alpha} 
\nonumber\\
=&
\frac{1}{2} \sum_{n=1}^{+\infty} \frac{(-1)^{n}}{(n \alpha)^2 e^{n \beta}}  
\left[e^{n \alpha}-e^{-n\alpha}-n\alpha (e^{n \alpha}+e^{-n\alpha})\right]
\nonumber\\
=&\frac{1}{2}\left[\frac{1}{\alpha^2}\sum_{n=1}^{+\infty}\frac{\left(-\frac{e^\alpha}{e^\beta}\right)^{n}}{n^2}-
\frac{1}{\alpha^2}\sum_{n=1}^{+\infty}\frac{\left(-\frac{e^{-\alpha}}{e^\beta}\right)^{n}}{n^2}-
\right.\nonumber\\
&\left.\hspace{2cm}
\frac{1}{\alpha}\sum_{n=1}^{+\infty}\frac{\left(-\frac{e^\alpha}{e^\beta}\right)^{n}}{n}-
\frac{1}{\alpha}\sum_{n=1}^{+\infty}\frac{\left(-\frac{e^{-\alpha}}{e^\beta}\right)^{n}}{n}
\right]\nonumber\\
=&\frac{1}{2\alpha^2}\left[{\rm Li}_2\left(-\frac{e^\alpha}{e^\beta}\right)-
{\rm Li}_2\left(-\frac{e^{-\alpha}}{e^\beta}\right)+
\right.\\ 
& \left. \hspace{2cm}
\alpha \ln\left(1+\frac{e^\alpha}{e^\beta}\right)+
\alpha \ln\left(1+\frac{e^{-\alpha}}{e^\beta}\right)\right]
\nonumber\,,
\end{align}
where ${\rm Li}_2(z)$ is the polylog function, with $|z|<1$ in this case, so that the power series converges.

\subsection{${\cal A}^\mu$ in the fluid rest system}
The transformation between the heavy quark rest system and the rest system of the fluid cell is given by
\begin{equation}
\left(\begin{array}{c}
{\cal A}_{\rm fluid}^0\\
\vec{\cal A}_{\rm fluid} \end{array}\right)=
\left(\begin{array}{cc}
u^0 & -\langle\vec{u} \\
-\vec{u}\rangle & 1 + \frac{\vec{u}\rangle \langle\vec{u}}{1+u^0} 
\end{array}\right)\cdot
\left(\begin{array}{c}
-A_{\rm rest}^0\\
-A_{\rm rest}^v \hat{u} 
\end{array}\right)
\end{equation}
where $u\equiv (u^0,\vec{u})$ is the fluid 4-velocity measured in the $c$-quark rest frame, that is 
$u=\frac{1}{m_Q}(E_p,-\vec{p})$. We thus get
\begin{align}
\vec{\cal A}_{\rm fluid}=&A_{\rm rest}^0 \vec{u} - A_{\rm rest}^v 
\underbrace{\left(1+\frac{\|\vec{u}\|^2}{1+u^0} \right)}_{=\frac{1+u^0+\|\vec{u}\|^2}{1+u^0}=u^0} \hat{u}
\nonumber \\ =&
\left(A_{\rm rest}^0 \|\vec{u}\| - A_{\rm rest}^v u^0  \right) \hat{u}
\nonumber\\
=&\left(A_{\rm rest}^v \frac{E_p}{m_Q}-A_{\rm rest}^0 \frac{\|\vec{p}\|}{m_Q} \right) \hat{p}\,.
\end{align}
Thus we find for the drag force in the fluid rest system
\begin{equation}
\vec{A}=\frac{m}{E_p}\vec{\cal A}_{\rm fluid}=
\left(A_{\rm rest}^v - \beta A_{\rm rest}^0\right) \hat{p}\,.
\end{equation}
with $\beta=\frac{p}{E_p}$. In Fig. \ref{fig_drag_C_et_E}, we present the drag force for the models C (top) and model E (bottom) of  \cite{Gossiaux:2008jv}.
For model C  $\alpha_s =\alpha_s(2\pi T)$ and the IR regulator $\mu^2 = 0.15 m_D^2$ in the propagator, while for model E, $\alpha_s =\alpha_{\rm eff}(t)$ and $\mu^2=0.2 \tilde{m}_D^2$ \cite{Gossiaux:2008jv}. The various curves correspond to $m_q=0$, $m_q=T$, $m_q=2T$ and $m_q=3T$.
One sees that giving a finite mass to the light quark leads to the reduction of the drag force for a given temperature.

\end{document}